\documentclass[10pt,letterpaper,showpacs]{article}
\pdfoutput=1

\usepackage{jheppub}
\usepackage{amsmath,amsfonts,amssymb,color,latexsym,theorem,mathrsfs,comment,color,dsfont}
\usepackage{tikz}
\usetikzlibrary{matrix}
\usetikzlibrary{positioning}
\usetikzlibrary{intersections,calc}
\usetikzlibrary{arrows.meta}
\usetikzlibrary{shapes,arrows}

\usepackage[english]{babel}
\usepackage[utf8]{inputenc}
\usepackage{slashed}

\newcommand{\ma}[1]{\mbox{$\mathcal{#1}$}}
\newcommand{\mas}[1]{\mbox{$\mathscr{#1}$}}
\newcommand{\D}{{\rm d}}
\newcommand{\ti}{\tilde}
\newcommand{\we}{\wedge}

\title{Supersymmetry of the Robinson-Trautman solution}

\author{
Masato Nozawa
}
\affiliation{
Department of General Education, Faculty of Engineering, 
Osaka Institute of Technology, 5-16-1, Omiya, Asahi-ku, 
Osaka, Osaka 535-8585, Japan.
}
\emailAdd{masato.nozawa@oit.ac.jp}

\abstract{
The Robinson-Trautman solution in the Einstein-Maxwell-$\Lambda$ system admits a shear-free and twist-free null geodesic congruence with a nonvanishing expansion. Restricting to the case where the Maxwell field is aligned, i.e., the spacetime is algebraically special, we provide an exhaustive classification of supersymmetric Robinson-Trautman spacetimes in the four-dimensional ${\cal N}=2$ gauged supergravity. The differential constraints that arise from the integrability conditions of the Killing spinor equation enable us to systematically reconstruct the metric. We derive the explicit form of the Killing spinor either by  directly integrating the Killing spinor equation or by casting the solution into the canonical form of supersymmetric solutions given in hep-th/0307022. In any case, the supersymmetric Robinson-Trautman solution generically exhibits a naked singularity.}


\begin{document}
\maketitle
\flushbottom

\section{Introduction}
\label{intro}

In recent years, gauged supergravity theories have significantly advanced our understanding of string theory and M-theory. 
The scope of gauged supergravities is very extensive, 
 ranging from the exploration of black hole physics to flux compactifications.  
Since the gauged supergravities accommodate  anti-de Sitter (AdS) spacetimes, 
they offer valuable testbeds for addressing quantum gravity and the behavior of strongly coupled field theories via holographic duality. Nevertheless, the exact asymptotically AdS solutions are hardly available due to the inapplicability of solution-generating methods~\cite{Klemm:2015uba}.

Among the gravitational solutions in supergravity, bosonic backgrounds preserving supersymmetry have played a pivotal role, since they are less susceptible to quantum corrections. The supersymmetric solutions possess a set of Killing spinors satisfying the first-order differential equations, which considerably constrain a possible form of spacetime metrics and fluxes. Consequently,  the Killing spinors facilitate the construction of exact solutions, providing crucial insights into the geometry and topology of the underlying spacetime.  Indeed, the Killing spinors prescribe the preferred $G$-structures, which yield a number of algebraic and differential relations for the bilinear tensors constructed out of the Killing spinors \cite{Gauntlett:2001ur,Gauntlett:2002sc}. The classification scheme based on the bilinear tensors, originally propounded in \cite{Gauntlett:2002nw}, enables us to cast the permissible metrics and fluxes into the canonical form. This prescription has been successfully applied to a variety of other supergravity theories in diverse dimensions~\cite{Gauntlett:2003fk,Gutowski:2004yv,Gutowski:2005id,Bellorin:2006yr,Bellorin:2007yp,Gutowski:2003rg,Gauntlett:2002fz,Gauntlett:2003wb,Caldarelli:2003pb,Bellorin:2005zc,Meessen:2010fh,Meessen:2012sr,Nozawa:2010rf}. 
Another algorithm for classifying supersymmetric solutions is the spinorial geometry, wherein the spinors are expressed in terms of form-fields and  subsequently brought into the preferred representative of their orbit~\cite{Gran:2005wn,Gran:2005ct,Cacciatori:2007vn,Gutowski:2007ai,Grover:2008ih,Cacciatori:2008ek,Klemm:2009uw,Klemm:2010mc}. 
For further insights into these developments, we direct readers to recent review articles \cite{Maeda:2011sh,Gran:2018ijr}.

In four-dimensional $\ma N=2$ ungauged supergravity, Tod made a pioneering contribution  \cite{Tod:1983pm}, prior to the above series of works, by obtaining a comprehensive  list of supersymmetric solutions utilizing the Newman-Penrose formalism \cite{Newman:1961qr}. In the case where the Killing vector built out of the Killing spinor is timelike, the solution reduces to the Israel-Perjes-Wilson family \cite{Perjes:1971gv,Israel:1972vx} which is completely specified by a complex harmonic function on the three-dimensional base space $\mathbb E^3$.  In four-dimensional $\ma N=2$ gauged supergravity, 
the systematic classification of supersymmetric solutions was undertaken  in \cite{Caldarelli:2003pb} and later scrutinized  in detail by \cite{Cacciatori:2007vn,Cacciatori:2004rt}.
It turns out that the solution belonging to the timelike class obeys a set of differential equations on a curved three-dimensional base space.\footnote{The holonomy of this space is reduced with respect to the torsionfull connection \cite{Klemm:2015mga}.}  In contrast to the ungauged case,  these differential equations remain nonlinear. This is a primary obstacle to obtaining exhaustive list of supersymmetric asymptotically AdS solutions. It therefore follows that the physical properties and the entire  landscape of supersymmetric solutions remain elusive in the gauged case, notwithstanding the overarching motivations such as holography and the attractor mechanism.

A first trailblazing work concerning supersymmetric static solutions to $\ma N=2$ gauged supergravity was carried out by 
Romans~\cite{Romans:1991nq} (see also \cite{Sabra:1999ux,Chamseddine:2000bk}). 
As demonstrated in~\cite{Romans:1991nq}, the Killing spinor equation is essentially dependent on a single variable for the Reissner-Nordstr\"om-AdS family, for which the direct integration of the Killing spinor equation is possible.   In the rotating case, several  authors have  investigated  the integrability conditions for the Killing spinor \cite{Caldarelli:1998hg,AlonsoAlberca:2000cs}. Since the integrability conditions are merely necessary conditions for supersymmetry \cite{vanNieuwenhuizen:1983wu}, it remains uncertain whether the configurations satisfying the integrability conditions in  \cite{Caldarelli:1998hg,AlonsoAlberca:2000cs} indeed admit a Killing spinor. It has been widely recognized as an unfeasible task to solve the Killing spinor equation directly, since equations depend nontrivially both on radial and angular coordinates
(see \cite{Lemos:2000wp} for an exceptional instance). 
Nevertheless, in \cite{Klemm:2013eca},  Klemm and the present author were able to demonstrate the existence of the Killing spinor for the Pleba\'nski-Demia\'nski solution~\cite{Plebanski:1976gy}, which describes the most general Petrov-D spacetime with an aligned non-null electromagnetic field and 
corresponds to a rotating, charged, and uniformly accelerating mass. The Pleba\'nski-Demia\'nski solution harbors two-commuting Killing vectors and encompasses the Reissner-Nordstr\"om-AdS, Kerr-Newman-AdS, and the AdS C-metric as special cases. 
The basic strategy  in \cite{Klemm:2013eca} is to implement the coordinate transformation 
in such a way that  the solution fits into the canonical form in \cite{Caldarelli:2003pb}, where the explicit solution to the Killing spinor equation is given, 
provided that the underlying nonlinear set of equations is solved. 
The technique  outlined in \cite{Klemm:2013eca} has seen broad application in finding rotating solutions in other supergravity theories, 
e.g., \cite{Gnecchi:2013mja,Nozawa:2015qea,Nozawa:2017yfl,Hristov:2019mqp,Ferrero:2020twa}.
In this paper, we delve  deeply into the supersymmetric Robinson-Trautman class of solutions which represents another avenue for generalizing the Reissner-Nordstr\"om-AdS family.

The Robinson-Trautman solution describes a radiating spacetime admitting an expanding congruence of null geodesics which is shear-free and twist-free \cite{RT,RT2}. In general, the Robinson-Trautman solution is devoid of Killing symmetry. The vacuum Robinson-Trautman solution is specified by a solution to the fourth-order nonlinear partial differential equation, which 
is identified as the Calabi flow~\cite{Tod}. Foster and Newman delved into the linear perturbation \cite{FN} 
and demonstrated that  perturbations decay exponentially at large retarded time, implying that  the spacetime eventually settles down to the Schwarzschild solution. Luk\'acs et al. demonstrated the validity of this argument at  the nonlinear level by exploiting the Lyapunov functional \cite{Lukacs:1983hr}. The global solution was addressed extensively by numerous authors \cite{Rendall,Singleton,Chrusciel:1991vxx,Chrusciel:1992rv,Chrusciel:1992tj}. Their work shows that the Robinson-Trautman solution exists globally for generic, arbitrarily strong, and smooth initial data, providing a conclusive proof that the solution eventually converges to the Schwarzschild metric. However, we must note that the existence of radiation implies an alluring physical consequence that the 
extension across the event horizon is available only with a finite degree of differentiability. 
Other physical properties were explored by many authors, including the asymptotic solutions
\cite{Vandyck,Vandyck2,Schmidt}, the B\"acklund transformation \cite{Glass,Hoenselaers,Hoenselaers2}, 
the Petrov types \cite{Podolsky:2016sff}, the cosmic no-hair conjecture \cite{Bicak:1995vc,Bicak:1997ne} and the holography  \cite{BernardideFreitas:2014eoi,Bakas:2014kfa,Adami:2024mtu}.

The charged Robinson-Trautman solution was constructed via the spin coefficient method~\cite{Newman:1969nvq} as well as 
the complexification technique \cite{LN}. However, its physical properties have been less explored. 
In this paper, the supersymmetry of the charged  Robinson-Trautman solution is worked out. 
Since the charged Robinson-Trautman solution is dynamical and specified by a set of differential equations, 
one might expect that there appear solutions endowed with richer physical structures than static solutions. 
This expectation motivates our study.

We lay out the present paper as follows. 
In the next section, we review the Robinson-Trautman solution in the Einstein-Maxwell-$\Lambda$ theory. 
In section~\ref{sec:integrability}, we derive the necessary conditions for supersymmetry  
from the integrability conditions of the Killing spinor equation. 
Section \ref{sec:classification} consists of the main results of our paper. 
By classifying the realizable supersymmetric solutions in a systematic fashion, 
we construct the explicit metric and gauge field, by integrating the integrability conditions. 
We discuss some physical aspects of each solution. 
The final conclusion is described in section \ref{sec:final}. 
Appendix~\ref{sec:g=0} discusses the ungauged Robinson-Trautman solution. 

\section{Charged Robinson-Trautman solution}

The action of the  Einstein-Maxwell-$\Lambda $ system is 
\begin{align}
\label{Lag}
S=\frac 1{16\pi G} \int \D ^4 x \sqrt{-g} \left(R-2\Lambda -F_{\mu\nu}F^{\mu\nu}\right)\,, 
\end{align}
where $F=\D A$ is the Faraday tensor and $\Lambda =-3g^2<0$. Here $g~(>0)$ is the inverse of the AdS radius. 
This is the bosonic part of the $\ma N=2$ minimal gauged supergravity in four dimensions. 
The field equations are 
\begin{align}
\label{Eineq}
R_{\mu\nu}=2 \left(F_{\mu\rho}F_\nu{}^\rho -\frac 14 g_{\mu\nu}F_{\rho\sigma}F^{\rho\sigma}\right)+\Lambda g_{\mu\nu} \,, 
\end{align}
and 
\begin{align}
\label{Maxwell}
\D F=0 \,, \qquad \D \star F= 0 \,. 
\end{align}
Here the star
denotes the Hodge dual operation 
 $(\star F)_{\mu\nu}=\frac 12 \epsilon_{\mu\nu\rho\sigma}F^{\rho\sigma}$.

The Robinson-Trautman solution describes a radiating spacetime that 
admits a twist-free and shear-free null geodesic congruence $\boldsymbol l$ with a nonvanishing expansion \cite{RT,RT}. 
In the $\Lambda$-vacuum case, the theorem of Goldberg-Sachs \cite{GS} allows us to find that the 
spacetime is algebraically special. The algebraically special feature is maintained in the Einstein-Maxwell-$\Lambda$ case, provided that 
the null direction $\boldsymbol l$ corresponds to the eigenvector of the Maxwell field, on which we shall focus in this paper.   In the aligned case, 
the charged  Robinson-Trautman solution reads \cite{Stephani:2003tm,Kozameh:2006hk,Griffiths:2009dfa}
\begin{align}
\label{metric}
\D s^2 &=-2 \D u \D r-H (u,r,z,\bar z)\D u^2 +2 r^2 P(u,z,\bar z)^{-2} \D z \D \bar z \,,
\end{align}
with 
\begin{align}
F=&-\frac{Q+\bar Q}{2r^2}\D u \we \D r 
+\frac 12 P^{-2} \left(Q-\bar Q\right)\D z \we \D \bar z \notag \\& 
+\D u \we \left[\left(\phi_2+\frac{\partial Q}{2r}\right)\D z
+\left(\bar \phi_2+\frac{\bar\partial \bar Q}{2r}\right)\D \bar z
\right] \,, 
\label{F}
\end{align}
where 
\begin{align}
\label{}
H (u,r,z,\bar z)&=K(u,z,\bar z)-\frac{2M(u,z,\bar z)}{r}+\frac{|Q(u, z)|^2}{r^2}
-2 r \big(\ln P(u,z,\bar z)\big)_u+g^2 r^2  \,. 
\end{align}
Here, 
we have introduced the abbreviation $\partial =\partial/\partial z$, 
$\bar \partial =\partial/\partial \bar z$ and 
$P_u =\partial P/\partial u$. The function $K(u, z, \bar z)$ is defined
in terms of $P(u,z,\bar z)$ as 
\begin{align}
\label{}
K(u, z, \bar z)\equiv 2P(u,z,\bar z)^2 \partial \bar \partial \ln P(u,z,\bar z)\,, 
\end{align}
which denotes the Gauss curvature of the (possibly $u$-dependent) two-dimensional 
space
\begin{align}
\label{2Dmetric}
\D s_2^2=g_{ij}^{(2)}(u, x^i)\D x^i \D x^j =\frac{2}{P^2}\D z \D \bar z \,, \qquad x^i=(z,\bar z)\,. 
\end{align}
It follows that the solution is specified by four $r$-independent functions $M=M(u,z,\bar z)$, 
$P=P(u,z,\bar z)$ ($>0$), $Q=Q(u,z)$ and $\phi_2=\phi_2 (u,z,\bar z)$. 
The functions $M$ and $P$ are real, whereas $Q$ and $\phi_2$ are complex. Note that $Q$ is holomorphic in the complex coordinate $z$. 
For the satisfaction of field equations (\ref{Eineq}) and (\ref{Maxwell}), 
these functions must obey the following set of partial differential equations
\begin{align}
\label{}
\partial M=&\,-2 \bar Q \phi_2 \,, \label{EOM1} \\
Q\bar Q_u-\bar Q Q_u=&\,2{P^2}(\bar \phi_2 \partial Q-\phi_2 \bar \partial \bar Q)\,,\label{EOM2} \\ 
\Delta_2 K =&\, 4M_u-12M (\ln P)_u+8P^2|\phi_2|^2
\label{EOM3} \,,\\
\bar \partial \phi_2 =&\,-\frac 12 (QP^{-2})_u\label{EOM4} \,. 
\end{align}
Equation (\ref{EOM2}) corresponds to the integrability condition $(\partial\bar\partial-\bar\partial\partial )M=0$ of (\ref{EOM1}) when equation (\ref{EOM4}) is fulfilled. 
Equation (\ref{EOM4}) ensures the existence of the gauge potential 
$A$ satisfying  $F=\D A$ as 
\begin{align}
\label{Apot}
A=- \frac{Q+\bar Q}{2r} \D u +\left(\int \phi_2 \D u\right)\D z
+\left(\int \bar \phi_2 \D u\right)\D \bar z \,.
\end{align}
The curvature invariant reads
\begin{align}
\label{}
R_{\mu\nu\rho\sigma}R^{\mu\nu\rho\sigma}=24g^4+\frac{48}{r^6}\left(M-\frac{|Q|^2}{r}\right)^2+\frac{8|Q|^4}{r^8}\,. 
\end{align}
It follows that $r=0$ is the spacetime curvature singularity except for $M=Q=0$.

In the coordinate system (\ref{metric}), the nontwisting shear-free null geodesics 
are generated by $\boldsymbol l=\partial/\partial r$, i.e.,  $r$ is an affine parameter of the null geodesics. 
Taking the Newman-Penrose null frame (we follow the notations in \cite{Stephani:2003tm})
\begin{align}
\label{NPframe}
\boldsymbol l= \frac{\partial}{\partial r}\,, \qquad 
\boldsymbol n=\frac{\partial}{\partial u} -\frac 12  H\frac{\partial}{\partial r} \,, \qquad 
\boldsymbol m= \frac P{r}\frac{\partial}{\partial z}\,, 
\end{align}
the metric reads 
\begin{align}
\label{}
g_{\mu\nu}=-2 l_{(\mu} n_{\nu)} + 2m _{(\mu}\bar m_{\nu)} \,. 
\end{align}
The Maxwell field is aligned with respect to $\boldsymbol l$, in the sense that 
\begin{subequations}
\begin{align}
\label{}
\Phi_0 \equiv&\, F_{\mu\nu}l^\mu m^\nu=0 \,, 
\\
\Phi_1 \equiv&\, \frac 12 F_{\mu\nu}(l^\mu n^\nu+\bar m^\mu m^\nu) =\frac{\bar Q}{2r^2}\,, 
\\
\Phi_2 \equiv&\, F_{\mu\nu}\bar m^\mu n^\nu=-\frac{P}{2r^2}\left(\bar \partial \bar Q+2r \bar \phi_2\right)\,.
\end{align}
\end{subequations}
In this aligned case, the Goldberg-Sachs' theorem is straightforwardly extended to the Einstein-Maxwell-$\Lambda$ system to conclude that 
the solution is algebraically special
\begin{align}
\label{}
\Psi_0\equiv &\, C_{\mu\nu\rho\sigma}l^\mu m^\nu l^\rho m^\sigma =0\,, \qquad 
\Psi_1\equiv \, C_{\mu\nu\rho\sigma}l^\mu n^\nu l^\rho m^\sigma =0\,.
\end{align}
It turns out that the Petrov type is at least degenerate to type-II. 
The nonvanishing Weyl scalars are given by
\begin{align}
\label{}
 \Psi_2\equiv\,& C_{\mu\nu\rho\sigma}l^\mu m^\nu \bar m ^\rho n^\sigma
 =-\frac{M}{r^3}+\frac{|Q|^2}{r^4} \,,  \\
\Psi_3\equiv\,&\, C_{\mu\nu\rho\sigma}n^\mu l^\nu n^\rho \bar m^\sigma=
 -\frac {P\bar \partial K}{2r^2}
-\frac{3PQ\bar \phi_2}{r^3}-\frac{PQ \bar \partial \bar Q}{r^4} \,, \\
\Psi_4\equiv\, & C_{\mu\nu\rho\sigma}n^\mu \bar m^\nu n^\rho \bar m^\sigma=
\bar \partial \left(-\frac{P^2 (\bar\partial \ln P)_u}{r}+\frac{P^2 \bar \partial K}{2r^2}+\frac{2QP^2 \bar \phi_2}{r^3}
+\frac{P^2 Q \bar \partial \bar Q}{2r^4}\right)
\,.
\end{align}
From the behavior of these curvature scalars, one deduces that $M$ and $Q$ play the 
role of mass and (complex) electromagnetic charge, respectively. 

The above null tetrad frame (\ref{NPframe}) is parallelly propagated along $\boldsymbol l$ as 
\begin{align}
\label{}
l^\nu \nabla_\nu l^\mu = l^\nu \nabla_\nu n^\mu = l^\nu \nabla_\nu m^\mu = 0\,.
\end{align}
It follows that the each Weyl scalar measures the Weyl curvature component in a basis that is 
parallelly propagated along to the null geodesics. Thus, the $r=0$ corresponds to the 
the parallelly propagated (p.p.)  curvature singularity even when $M=Q=0$.

It is worth mentioning that the line element (\ref{metric}) and the Maxwell field (\ref{F}) are 
preserved under the reparameterization
\begin{align}
u\to\,& U(u) \,, & r\to \,&r U_u ^{-1}\,, & z\to\, &\zeta (z)\,, &
P\to\, & P |\partial \zeta| U_u^{-1} \,,   \notag \\
K\to &\,U_u^{-2} K\,,  & M\to &\, M U_u^{-3} \,, &
Q \to\,  & Q U_u^{-2}\,, & \phi_2\to \,& U_u^{-1} (\partial \zeta)^{-1} \phi_2 \,,
\label{repara}
\end{align} 
where $U=U(u)$ is real and $\zeta=\zeta(z)$ is holomorphic in $z$. 
Furthermore, the Einstein equations (\ref{Eineq}), 
the Maxwell field equations and the Bianchi identity 
(\ref{Maxwell}) are invariant under the ${\rm U}(1)$ electromagnetic duality transformations
\begin{align}
\label{duality}
F+i \star F \to e^{i\gamma }(F+i\star F)\,, 
\end{align}
where $\gamma$ is a constant. Here we take the orientation as 
$\boldsymbol \epsilon=-i r^2 P^{-2}\D u \we \D r\we \D z \we \D \bar z$. 
Under the electromagnetic duality,  $Q$ and $\phi_2$ transform as
\begin{align}
\label{duality2}
Q\to e^{i\gamma} Q\,, \qquad \phi_2 \to e^{i\gamma} \phi_2 \,. 
\end{align}
It should be kept in mind that the electromagnetic duality is the 
invariance of equations of motion.  This transformation
retains neither the Lagrangian nor the Killing spinor equation, as we will 
see in the next section.

Since the general Robinson-Trautman family (\ref{metric}) is fairly general, let us now consider 
its subclasses to capture the physical meaning.

\subsection{AdS}
\label{sec:AdS} 

To begin with, 
it is instructive to see how the background AdS spacetime is embedded into the Robinson-Trautman family. 
The maximally symmetric AdS space is achieved if and only if all the Weyl scalars and the Maxwell scalars vanish, implying  
$M=Q=\phi_2=0$ with 
\begin{align}
\label{AdS_Peq}
 \partial K=0 \,, \qquad 
(P^{-1}  \partial^2 P)_u=0 \,. 
\end{align}
Upon integration, one finds a real function $K_0=K_0(u)$ and a complex function $h_0=h_0(z,\bar z)$ with 
\begin{align}
\label{AdS_Peq2}
K=K_0(u)\,, \qquad 
\partial ^2 P=h_0 (z,\bar z) P\,.
\end{align}
Taking $\bar z$ derivative of the second equation, the compatibility of 
above equations leads to $\bar \partial h_0=0$, i.e., $h_0$ is a holomorphic function of $z$. 
Using the reparametrization freedom (\ref{repara}), one can set $K_0(u)=k=\{0, \pm 1\}$ without losing any generality.
Since the first equation of (\ref{AdS_Peq2}) is the $u$-dependent Liouville's equation for $P$, 
its general solution in the simply connected region is given by
\begin{align}
\label{}
P(u,z,\bar z)= \frac{1+\frac 12 k |h_1 |^2}{(\partial h_1 \bar \partial \bar h_1 )^{1/2} } \,, 
\end{align}
where $h_1=h_1(u,z)$ admits at most simple poles \cite{Henrici}. 
Inserting this into the second equation of (\ref{AdS_Peq2}), one finds that $h_1(u,z)$ must satisfy
\begin{align}
\label{}
 \{h_1(u,z) , z \}=-2 h_0(z)\,,
\end{align}
where the right hand side of this equation denotes the Schwarzian derivative
\begin{align}
\label{}
\{h_1(u,z) , z \} \equiv 
\frac{\partial^3 h_1}{\partial h_1}-\frac 32 \left(\frac{\partial^2 h_1}{\partial h_1}\right)^2 \,. 
\end{align}
This means that $h_1$ is given by $h_1(u,z)=f_1(u,z)/f_2(u,z)$, where $f_1$ and $f_2$ are linearly independent holomorphic solutions to 
the differential equation
\begin{align}
\label{fieq}
\partial^2 f_i(u,z)  -h_0(z) f_i (u,z)=0 \,.  
\end{align}

One possible way to realize AdS is to choose $h_1$ to be independent of $u$, 
for which the reparameterization freedom (\ref{repara}) enables us to set
\begin{align}
\label{}
h_0(z)=0 \qquad 
h_1=z \,.
\end{align}
This yields the AdS metric in the conventional  form
\begin{align}
\label{AdSmetric}
\D s^2 = -\left(k+g^2r^2\right)\D u^2-2\D u \D r+r^2 \D \Sigma_k^2 \,, 
\end{align}
where $\D \Sigma_k^2$ is the two-dimensional metric of constant Gauss curvature $k$, 
\begin{align}
\label{Pk}
\D \Sigma_k^2=2 P_k^{-2} \D z \D \bar z \,, \qquad 
P_k=1+\frac k2 z\bar z\,. 
\end{align} 
Alternatively, a coordinate transformation $z=\sqrt{2/k}\tan (\sqrt k \theta/2)e^{i\varphi}$
leads to a manifestly axisymmetric form 
\begin{align}
\label{dSigmasq}
\D \Sigma_k^2= \D \theta^2 +\left(\frac 1{\sqrt k}\sin (\sqrt k \theta)\right)^2 \D \varphi^2 \,.
\end{align}

\subsection{Vacuum case} 

In the vacuum case $Q=\phi_2=0$, equation (\ref{EOM1}) implies $M=M(u)$, while 
the reparametrization freedom (\ref{repara}) enables us to set $M$ to be a constant. 
Then, the vacuum Robinson-Trautman equation (\ref{EOM3}) amounts to the Calabi-flow equation for the two-dimensional 
metric (\ref{2Dmetric}) as~\cite{Tod}
\begin{align}
\label{}
\frac{\partial }{\partial t}g_{z\bar z}^{(2)} =\partial \bar \partial R^{(2)}\,, 
\end{align}
where $t=u/(6M)$ is a time coordinate of the flow, $g^{(2)}_{z\bar z}$ is the 
two-dimensional (K\"ahler) metric and $R^{(2)}=2K$ is its scalar curvature. 
This is a parabolic, nonlinear partial differential equation of fourth order, which resembles considerably the heat flow diffusion equation. 
This flow equation is well-suited for analyzing the global behavior of the solution, since it defines an appropriate functional 
that is positive-definite and monotonically decreasing along the flow~\cite{Lukacs:1983hr}. 

In the charged case, on the other hand, it appears that the generalized Robinson-Trautman equation
(\ref{EOM3}) cannot be recast into the first-order flow equation. This is a primary adversity when trying to 
explore the global behavior of the charged Robinson-Trautman solution. See \cite{Lun:1994up,Kozameh:2007yf,Chrusciel:2021pnv}
for related discussions.

\section{Integrability conditions for supersymmetry}
\label{sec:integrability}

The bosonic part of the minimal $\ma N=2$ gauged supergravity (\ref{Lag}) consists of vielbein $e^a{}_\mu$ and 
the ${\rm U}(1)$ gauge field $A_\mu$ \cite{Freedman:1976aw}. This four-dimensional theory can be embedded into eleven-dimensional supergravity on the deformed seven sphere \cite{Chamblin:1999tk,Gauntlett:2006ai}.

The Killing spinor equation for minimal $\ma N=2$ gauged supergravity is given by \cite{Freedman:1976aw}
\begin{align}
\label{Killingspinor}
\hat \nabla_\mu \epsilon \equiv \left(\nabla_\mu +\frac{i}4 F_{\nu\rho }
\gamma^{\nu\rho }\gamma_\mu +\frac{1}{2}g\gamma_\mu -i g A_\mu \right)\epsilon  =0 \,,
\end{align}
where $\nabla_\mu=\partial_\mu+\frac 14\omega_{\mu ab}\gamma^{ab}$ is a covariant derivative
acting on the spinor, where $\omega_{\mu ab}=e_{a\nu}\nabla_\mu e_b{}^\nu$ denotes the spin connection.
Gamma matrices satisfy  $\gamma_{(\mu}\gamma_{\nu)}=g_{\mu\nu}$ and $\gamma_{\mu\nu}=\gamma_{[\mu}\gamma_{\nu]}$. 
Under the gauge transformation $A_\mu \to A_\mu +\nabla_\mu \chi$, 
the Killing spinor equation remains invariant, provided that the Killing spinor transforms as
\begin{align}
\label{gaugetr}
\epsilon \to e^{i g \chi} \epsilon \,. 
\end{align}
To rephrase, the Killing spinor possesses a gauge charge.

Taking the supercovariant derivative of (\ref{Killingspinor}) and antisymmetrizing indices, 
we obtain the integrability conditions for the Killing spinor equation
\begin{align}
\label{supercurvature}
\ma R_{\mu\nu} \epsilon =0 \,, \qquad  \ma R_{\mu\nu}\equiv 
[\hat \nabla_\mu , \hat \nabla_\nu]\,, 
\end{align}
where $\ma R_{\mu\nu} $ denotes the supercurvature. The viable structure of 
supercurvature in the context of positive energy theorem was worked out in \cite{Nozawa:2013maa}. 
For the existence of the nontrivial solutions to 
the algebraic equations $\ma R_{\mu\nu} \epsilon=0$, 
the following integrability conditions must be satisfied 
\begin{align}
\label{intcond}
{\rm det}(\ma R_{\mu\nu})=0 \,. 
\end{align}
It is tedious  but  straightforward to compute  
${\rm det}(\ma R_{\mu\nu})=0$ for the Robinson-Trautman solution. For the calculation,  
we take the tetrad frame 
\begin{align}
\label{}
e^+ =\D r+\frac 12 H \D u \,, \qquad 
e^- = \D u \,, \qquad 
e^\bullet = r P^{-1} \D z \,, \qquad 
e^{\bar \bullet}= r P^{-1}\D \bar z \,, 
\end{align}
with 
\begin{align}
\label{}
g_{\mu\nu}= \eta_{ab} e^a{}_\mu e^b{}_\nu \,, \qquad  
\eta_{ab}= \left(\begin{array}{cccc}
 0 &- 1 & 0& 0   \\
 -1 & 0 & 0& 0   \\ 
 0&0& 0 & 1\\
 0&0& 1&0 
\end{array}\right)\,, 
\end{align}
and the following gamma matrix representation
\begin{align}
\label{}
\gamma_+ &=\left(
\begin{array}{cccc}
     0 & 0 & 0 & 0     \\
     0 & 0 & 0 & \sqrt 2     \\
 \sqrt 2 & 0 & 0 & 0     \\
   0 & 0 & 0 & 0    
\end{array}
\right)\,, \qquad 
\gamma_- =\left(
\begin{array}{cccc}
     0 & 0 & -\sqrt 2 & 0     \\
     0 & 0 & 0 & 0     \\
   0 & 0 & 0 & 0     \\
   0 & -\sqrt 2 & 0 & 0    
\end{array}
\right)\,, \notag \\
\gamma_\bullet &=\left(
\begin{array}{cccc}
     0 & 0 & 0 & 0     \\
     0 & 0 & -\sqrt 2 & 0     \\
   0 & 0 & 0 & 0     \\
 \sqrt 2 & 0 & 0 & 0    
\end{array}
\right)\,, \qquad 
\gamma_{\bar \bullet} =\left(
\begin{array}{cccc}
     0 & 0 & 0 & \sqrt 2     \\
     0 & 0 & 0 & 0     \\
   0 & -\sqrt 2 & 0 & 0     \\
   0 & 0 & 0 & 0    
\end{array}
\right)\,, \label{gammamat}
\end{align}
with $\gamma_5=\gamma_{+-\bullet \bar \bullet}={\rm diag}(-1,-1,1,1)$. 
We find that only the ($u,r$)-component of (\ref{intcond}) gives rise to nontrivial conditions, which are reduced to 
the following form
\begin{align}
\label{}
{\rm det}(\ma R_{ur})=\frac{\ma B_1}{r^{12}}
+\frac{\ma B_2-i \ma B_3}{r^{11}}+\frac{\ma B_4-i \ma B_5}{2r^{10}} \,, 
\end{align}
where each $\ma B_i$ is real and independent of $r$. Their explicit expressions read 
\begin{subequations}
\label{Bi}
\begin{align}
\label{}
\ma B_1\equiv &\, (M^2-|Q|^2 K)^2 +2 P^2 \left|M  \partial Q+2 |Q|^2 \phi_2\right|^2 +g^2 |Q|^4 (Q-\bar Q)^2 \,,
\\
\ma B_2\equiv &\, -P^4(|Q|^2P^{-4})_u (M^2-|Q|^2K)-2g^2 |Q|^2 (Q-\bar Q)^2 M
\notag \\ &
+P^2 \left\{(Q\partial K+2M \phi_2)(M\bar \partial \bar Q+2 |Q|^2\bar \phi_2)+{\rm c.c}\right\} \,,
\\
\ma B_3\equiv &\, g P^2 Q \bar \partial \bar Q(M\partial Q+2|Q|^2 \phi_2)+{\rm c.c} \,,
\\
\ma B_4\equiv &\,  P^2 |Q\partial K+2M \phi_2|^2+2|Q|^2 P^4 (QP^{-2})_u(\bar Q P^{-2})_u 
\notag \\
&+g^2 \left[ 2M^2 (Q-\bar Q)^2- P^2 |Q|^2 |\partial Q |^2\right]\,,  
\\
\ma B_5\equiv &\,g\Bigl[
\left\{P^2 \bar Q \bar\partial \bar Q(Q\partial K+2M \phi_2)+{\rm c.c}\right\}
+2M(Q-\bar Q)(Q\bar Q_u-\bar Q Q_u)
\Bigr]\,.
\end{align}
\end{subequations}
Here we have imposed (\ref{EOM1}) and (\ref{EOM2}). 
It turns out that the integrability conditions put the following five constraints
\begin{align}
\label{}
\ma B_i =0\,, \qquad i=1,...,5. 
\end{align}

Defining complex scalars $Z_i$ and real scalars $X_i$ ($i=1,...,4$) by
\begin{align}
\label{}
Z_1=&\, P(M\partial Q+2 |Q|^2 \phi_2 )\,, &
Z_2=&\,P(Q\partial K+2M\phi_2)\,,\notag \\
Z_3=&\,\bar Q P^2(QP^{-2})_u \,,&
Z_4=&\,g P\bar \partial \bar Q\,, 
\end{align}
and 
\begin{align}
\label{}
X_1=&\,M^2-|Q|^2K\,, & X_2=&\,ig (Q-\bar Q) \,,\notag \\
X_3=&\,P^4 (|Q|^2 P^{-4})_u \,, & X_4=&\,-i(Q\bar Q_u-\bar Q Q_u)\,,
\end{align}
the supersymmetry conditions are recast into a simpler form
\begin{subequations}
\label{BiZX}
\begin{align}
\ma B_1=&\, X_1^2+2|Z_1|^2-|Q|^4X_2^2 \,, \\
\ma B_2=&\,-X_3X_1+Z_2\bar Z_1+\bar Z_2 Z_1+2M|Q|^2X_2^2 \,, \\
\ma B_3=&\,QZ_4Z_1+\bar Q \bar Z_4\bar Z_1 \,, \\
\ma B_4=&\,|Z_2|^2+2|Z_3|^2-2M^2 X_2^2-|Q|^2|Z_4|^2 \,, \\
\ma B_5=&\,\bar QZ_4Z_2+Q\bar Z_4 \bar Z_2+2MX_2X_4\,. 
\end{align}
\end{subequations}
These scalars are  subjected to the constraints 
\begin{align}
\label{constraints1}
Z_3=&\,\frac 12 (X_3-iX_4) \,, \\
ig |Q|^2 X_4=&\, -(Z_1Z_4-\bar Z_1\bar Z_4)\,.
\label{constraints2}
\end{align}

Several issues are worthy of emphasis. First,  
the supersymmetry conditions give rise to differential restrictions, which depend 
in a fairly nontrivial way on $z$, $\bar z$ and $u$. 
This is in sharp contrast to the  supersymmetry conditions for the Pleba\'nski-Demia\'nski solution, 
where they are purely algebraic \cite{Klemm:2013eca}, because the Pleba\'nski-Demia\'nski solution does not involve arbitrary functions. 
Nevertheless, a vital observation is that these supersymmetry conditions (\ref{BiZX}) do not involve the second derivative of $K$. 
This means that the degree of the differential equation has decreased compared to equations of motion. 
As a matter of fact, even though it is a formidable task to solve equations of motion 
(\ref{EOM1})--(\ref{EOM4}) in full generality,  it is possible to find the exhaustive list 
of supersymmetric solutions, as we will demonstrate in the following. 
This feature epitomizes the restrictive nature of supersymmetry. 
With these remarks in mind, our remaining tasks are (1) to classify the possible cases systematically, (2) to obtain each explicit metric and (3) to check the existence of the Killing spinor. The significance of the last step (3)  cannot be overstated, since the first 
integrability conditions (\ref{intcond}) are merely the necessary conditions for the existence of the Killing spinor~\cite{vanNieuwenhuizen:1983wu}.

Second, the Killing spinor equation (\ref{Killingspinor}) is {\it not} invariant under the electromagnetic duality (\ref{duality}). The principal reason is that the duality transformation (\ref{duality}) acts on the complexified field strength $F_{\mu\nu}+i \star F_{\mu\nu}$, rather than the gauge potential $A_\mu$. 
Although the electromagnetic duality is the symmetry of the equations of motion, it does not maintain 
the Killing spinor equation, since the Killing spinor has the gauge charge (\ref{gaugetr}). 
The most unequivocal manifestation of this fact is provided by the supersymmetric cosmic dyon \cite{Romans:1991nq}, for which the electric charge makes no contribution to the supersymmetry condition. 
This solution will be encountered in the next section. 
The lack of invariance of the Killing spinor under the electromagnetic duality  is also made more compelling by the fact that  the Bogomol'ny bound in AdS--first advocated in  
\cite{Kostelecky:1995ei}--fails to admit the magnetic charge, 
from the viewpoints of the superalgebra \cite{Dibitetto:2010sp} and the positive energy theorem \cite{Nozawa:2014zia}.

\section{Classification}
\label{sec:classification}

We now turn to the main part of this work. 
We implement the systematic classification of explicit supersymmetric solutions 
belonging to the Robinson-Trautman family. In this section, we limit ourselves exclusively to the gauged case ($g\ne 0$).
The discussion of the ungauged case ($g=0$) is postponed to appendix \ref{sec:g=0}.

For the systematic classification of solutions, 
it is adequate  to perform the  polar decomposition $Z_i=R_i e^{i\Theta_i}$ ($i=1,..,4$) 
for the complex scalars, 
 where $R_i=R_i(u,z,\bar z)$ and 
$\Theta_i=\Theta_i(u,z,\bar z)$ are real functions. 
In terms of these variables, the supersymmetry conditions are reduced to 
\begin{subequations}
\label{BPSp}
\begin{align}
\label{B1p}
\ma B_1=&\, X_1^2+2R_1^2-4g^2 q^6\sin^2\alpha  \,, \\
\label{B2p}
\ma B_2=&\,-2X_1R_3\cos\Theta_3+2R_1 R_2\cos(\Theta_1-\Theta_2)+8g^2 Mq^4 \sin^2\alpha \,, \\
\label{B3p}
\ma B_3=&\,2q R_1R_4 \cos(\alpha+\Theta_1+\Theta_4) \,, \\
\label{B4p}
\ma B_4=&\,R_2^2+2R_3^2-q^2\left(R_4^2+8 g^2 M^2 \sin^2\alpha\right) \,, \\
\label{B5p}
\ma B_5=&\,2q\left[R_2R_4\cos(\alpha-\Theta_2-\Theta_4)+4gMR_3\sin\Theta_3 \sin\alpha\right]\,,
\end{align}
\end{subequations}
with
\begin{align}
\label{constp}
gq^2 R_3 \sin\Theta_3=R_1 R_4 \sin(\Theta_1+\Theta_4)
\end{align}
Here, we have also introduced real scalars 
$q(u,z,\bar z)$ and $\alpha (u,z,\bar z)$ by 
\begin{align}
\label{normphase}
q(u,z,\bar z)=\sqrt{Q\bar Q}\,, \qquad 
e^{i\alpha(u,z,\bar z)}=Q/q\,.
\end{align}

\begin{figure}
\begin{center}
\includegraphics[width=14cm]{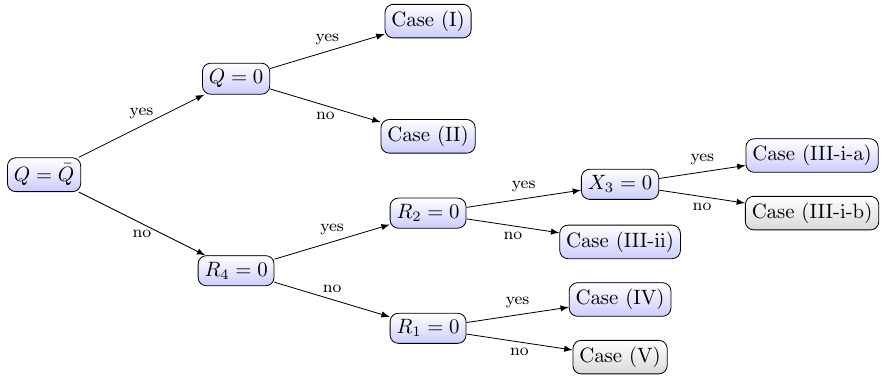}
\caption{Flowchart of classification for supersymmetric Robinson-Trautman solutions. 
The gray-shaded cases do not admit supersymmetric solutions.}
\label{fig:flowchart}
\end{center}
\end{figure}

We are now ready to proceed with the classification of supersymmetric Robinson-Trautman solutions. 
We first divide the analysis into $Q=\bar Q$ or not. 
In the case of the real charge function ($Q=\bar Q$),  we can consider two subclasses: the Case (I) $Q=0$ and Case (II) $Q\ne 0$. 
For $Q\ne \bar Q$, the inspection of (\ref{B3p}) allows us to 
categorize the solutions according to 
(III) $R_4=0$, (IV) $R_1=0$ ($R_4\ne0$) and 
(V) $\cos(\alpha+\Theta_1+\Theta_4)=0$ ($R_1R_4\ne 0$). 
See figure~\ref{fig:flowchart}  for the flowchart of classification.

\subsection{Case (I): $Q=0$}
\label{sec:Q0}

Let us first begin with the $Q=0$ case, for which $\ma B_1=0$ implies  $M=0$. Then,  
other conditions $\ma B_2=\ma B_3=\ma B_4=\ma B_5=0$ are trivially met. 
Now, the field equations imply
\begin{align}
\label{CaseIeq}
\Delta_2 K=8P^2 |\phi_2|^2 \,, \qquad 
\bar \partial \phi_2=0\,. 
\end{align}
The solution is therefore given by 
\begin{align}
\label{Case1sol}
\D s^2=&\,-2\D u \D r-\left[K-2r(\ln P)_u+g^2r^2\right]\D u^2 +\frac{2r^2}{P^2}\D z \D \bar z \,, \\
A=&\, -(\psi+\bar \psi)\D u  \,,
\end{align}
where we have written $\phi_2(u,z)=\partial \psi(u,z)$ and performed the gauge transformation 
$A\to A+\D \chi$, $\chi=-\int (\psi+\bar \psi)\D u$ compared to (\ref{Apot}). 
Since the gravitational Coulomb part vanishes, 
this solution belongs to Petrov type III, due to $\Psi_0=\Psi_1=\Psi_2=0$.
Up to this point, the source function $\psi(u,z)$ is completely arbitrary.

We next explore if this type III metric admits supersymmetry or not, by directly integrating the Killing spinor equation. 
As it turns out, the functions $P$ and $\psi $ are subjected to more constraints to preserve supersymmetry. 

The $r$-component of the Killing spinor is immediately integrated to give 
\begin{align}
\label{}
\epsilon =\Xi_r  \epsilon' \,, \qquad \Xi_r=\mathds{1}_4 -\frac 12 g r\gamma_+ \,,
\end{align}
where $\epsilon'$ is an $r$-independent spinor. 
The solution $\epsilon$ to the Killing spinor equation (\ref{Killingspinor}) satisfies $\Pi\epsilon=0$, where 
the projection operator $\Pi=\ma R_{ur}/P$ reads explicitly 
\begin{align}
\label{}
\Pi= \left(\frac{\bar \partial K}{4r^2}+\frac{i g\bar \phi_2}{2r}\right)\gamma_{+\bullet}
+\left(\frac{\partial K}{4r^2}+\frac{i g\phi_2}{2r}\right)\gamma_{+\bar\bullet}
-\frac{i  \phi_2}{2r^2}(\mathds 1_4+\gamma_5)\gamma_{\bar \bullet}
-\frac{i  \bar \phi_2}{2r^2}(\mathds 1_4-\gamma_5)\gamma_{\bullet}\,.
\end{align}
Under the projection $\Pi\epsilon=\Pi \Xi_r \epsilon'=0$, 
the remaining components of the Killing spinor equation are $r$-independent and are boiled down to
\begin{subequations}
\label{CaseI_partialep}
\begin{align}
\label{}
\left[\partial_u 
+\frac 12 \partial_u (\ln P)\gamma_{+-} +\frac 12 g\gamma_-+i g (\psi+\bar \psi)\mathds 1_4 +\frac 14 K g \gamma_+
\right. ~~~~~~~\notag \\
\left.
-\frac 12 P\left\{(\partial (\ln P)_u+i g \partial \psi) \gamma_{+\bar \bullet}
+(\bar \partial (\ln P)_u+i g \bar \partial \bar \psi)\gamma_{+\bullet}
\right\} 
\right]\epsilon'&\,= 0\,,\label{KSQ0u}
\\
\left[\partial -\frac{K}{4P}\gamma_{+\bullet}-\frac 12 \partial (\ln P)\gamma_{\bullet\bar\bullet}
+\frac 1{2P}\gamma_{-\bullet}-\frac i2 \partial \psi(\mathds 1_4+\gamma_5)\gamma_+ 
\right]\epsilon'&\,= 0\,,\label{KSQ0z}
\\
\left[\bar\partial -\frac{K}{4P}\gamma_{+\bar \bullet}+\frac 12 \partial (\ln P)\gamma_{\bullet\bar\bullet}
+\frac 1{2P}\gamma_{-\bar \bullet}-\frac i2 \bar \partial\bar \psi(\mathds 1_4-\gamma_5)\gamma_+ 
\right]\epsilon'&\,= 0\,.\label{KSQ0zb}
\end{align}
\end{subequations}
We now suppose $\partial \psi\ne 0$ and $ \partial K\ne 0$, since otherwise 
the solution would reduce to AdS, c.f. section \ref{sec:AdS}. 
Writing the spinor components corresponding to the 
 gamma matrix representation (\ref{gammamat}) as $\epsilon'=(\epsilon_1', \epsilon_2', \epsilon_3', \epsilon_4')^T$, 
the condition $\Pi\,\Xi_r \epsilon'=0$ projects out  half of spinor components as
\begin{align}
\label{epsilon32CaseI}
\epsilon_3'=\frac{i\bar \partial K}{2\sqrt 2 \bar \partial \bar \psi}\epsilon_1' \,, \qquad 
\epsilon_2'=\frac{i \partial K}{2\sqrt 2 \partial \psi}\epsilon_4' \,. 
\end{align}
Taking the derivative of $\Pi\, \Xi_r \epsilon'=0$, we have 
\begin{align}
\label{}
\mas R_{(\mu)}\equiv (\partial_\mu \Pi)\Xi_r \epsilon'+\Pi\,(\partial_\mu\Xi_r) \epsilon'+\Pi\,\Xi_r\partial_\mu \epsilon'=0\,, 
\end{align}
We shall refer to this equation as the second integrability conditions. 
Inspecting $\mas R_{(z)}=\mas R_{(\bar z)}=0$ and (\ref{CaseI_partialep}), 
we have the $2\times 2$ linear system $\ma M_1 \vec \epsilon^{\,\prime}=0$, where 
 $\vec \epsilon^{\,\prime}=(\epsilon_1', \epsilon_4')^T$ and 
\begin{align}
\label{}
\ma M_1\equiv \left(
\begin{array}{cc}
i \left(\sqrt 2 K\partial \psi-\frac{|\partial K|^2}{2\sqrt 2\bar \partial\bar \psi}\right)     & P\left(\partial^2 K-\frac{\partial^2 \psi\partial K}{\partial \psi}\right)    \\
P\left(\bar\partial^2 K-\frac{\bar\partial^2 \bar\psi\bar\partial K}{\bar\partial \bar\psi}\right)        &   i \left(\sqrt 2 K\bar \partial\bar \psi-\frac{|\partial K|^2}{2\sqrt 2\partial \psi}\right) 
\end{array}
\right)\,.
\end{align}
The nontrivial solutions to $\ma M_1\vec \epsilon^{\, \prime}=0$ exist only if ${\rm det}\ma M_1=0$. Here, we remark 
that the general solution to (\ref{CaseIeq}) is given by 
\begin{align}
\label{}
K(u,z,\bar z)=4 |\psi(u,z)|^2+\kappa (u,z)+\bar \kappa(u,\bar z) \,,
\end{align}
where $\kappa(u, z)$ is an arbitrary function independent of $\bar z$. 
Plugging this into ${\rm det}\ma M_1=0$, we obtain 
$\kappa(u, z)=4\kappa_0(u)\psi(u,z)+2|\kappa_0|^2+i \kappa_1(u)$, 
where $\kappa_0(u)$ is complex and $\kappa_1(u)$ is real.  
In this case, $K=4|\psi+\bar \kappa_0|^2$, to which $\kappa_1$ does not contribute.
$\bar\kappa_0$ can be set to zero 
by the redefinition $\psi \to \psi-\bar \kappa_0$ together with the gauge transformation
$A_\mu \to A_\mu +\nabla_\mu \chi$,  thus $K=4|\psi|^2>0$. 
Defining $P_* \equiv P/(2|\psi|)$, we obtain 
$K_*\equiv 2P_*^2\partial \bar\partial \ln P_*=1$. 
It follows that $2P_*^{-2}\D z \D \bar z$ is a $u$-dependent metric of unit two-sphere. 
By the  reasoning  in section \ref{sec:AdS}, we can thus set 
\begin{align}
\label{Pstar}
P=2|\psi|P_* \,, \qquad 
P_*=\frac{1+\frac 12 |h_1|^2}{|\partial h_1|}\,, \qquad 
h_1(u,z)=\frac{f_1(u,z)}{f_2(u,z)}\,, \qquad 
\partial^2 f_i =h_0(z) f_i\,.
\end{align}
This is the condition which cannot be captured by the first integrability conditions ${\rm det}(\ma R_{\mu\nu})=0$
for the Killing spinor.

The $u$ and $\bar z$ components of (\ref{KSQ0z}) and (\ref{KSQ0zb}) 
are integrated to give 
\begin{align}
\label{chi}
\epsilon_1'=\sqrt{\frac{\bar \psi}P}\bar \varepsilon _1(u, \bar z)\,, \qquad 
\epsilon_4'= \sqrt{\frac{\psi}{P}}\varepsilon _2(u, z)\,,
\end{align}
where $\varepsilon_i(u,z)~(i=1,2)$ are $u$-dependent holomorphic functions. 
The remaining components of (\ref{KSQ0z}) and (\ref{KSQ0zb})  yield
\begin{align}
\label{chiz}
\bar\partial \left(P_*^{-1}\bar\varepsilon_1\right)=-\frac{i}{\sqrt 2} P_*^{-2} \varepsilon_2 \,, \qquad 
\partial \left(P_*^{-1}\varepsilon_2\right)=-\frac{i}{\sqrt 2} P_*^{-2}\bar \varepsilon_1 \,,
\end{align}
implying $\partial^2 \varepsilon _i=h_0(z) \varepsilon_i$. The above relation
(\ref{chiz}) means that $\varepsilon_1=0$ implies $\varepsilon_2=0$ and vice versa, 
which  enforces both of $\varepsilon_i$ to be nonvanishing. 
The  $u$ and $\bar z$ components of (\ref{KSQ0u}) give
 \begin{align}
\label{psiCaseI}
 \psi =\frac 1{2ig}\left(\frac{(\varepsilon_1)_u}{\varepsilon_1}-\frac{(\varepsilon_2)_u}{\varepsilon_2}\right)\,,
\end{align}
and
\begin{align}
\label{chiuu}
\frac{(\varepsilon_1)_{uu}}{\varepsilon_1}=\frac{(\varepsilon_2)_{uu}}{\varepsilon_2} \,.
\end{align}
Under these conditions, all components of the Killing spinor equation are satisfied. 
Thus, the problem at hand is whether the functions $\varepsilon_i$ satisfying 
(\ref{chiz}), (\ref{chiuu}) and $\partial^2 \varepsilon _i=h_0(z) \varepsilon_i$ exist or not. 

To see this, let us expand $\varepsilon _i$ and $f_i$ in terms of $u$-independent holomorphic functions $\varsigma_i(z)$
satisfying $\partial^2 \varsigma _i=h_0(z) \varsigma_i$ as 
\begin{align}
\label{varepsilonex}
\varepsilon _i (u,z) =\sum_{j=1}^2 A_{ij}(u) \varsigma_j (z)\,, \qquad 
f _i (u,z) =\sum_{j=1}^2B_{ij}(u) \varsigma_j (z) \,.
\end{align}
Equation (\ref{chiuu}) then implies
\begin{align}
\label{Aijuu}
(A_{ij})_{uu}=a_0(u) A_{ij} \,, 
\end{align}
where $a_0$ is a function of $u$, whereas
equation  (\ref{chiz})  leads to 
\begin{align}
\label{Aij}
A_{22}=-\frac{i \bar A_{11}B_{12}\sqrt{\bar W_0}}{\sqrt{2W_0}\bar B_{21}}
=\frac{B_{12}}{B_{11}}A_{21} 
\,, \qquad 
A_{12}=\frac{B_{22}}{B_{21}}A_{11} \,.
\end{align}
Here we have defined $W_0 (u)\equiv (B_{11}B_{22}-B_{12}B_{21})\ma W(\ne 0)$, where  
$\ma W=\varsigma_2 \partial \varsigma_1-\varsigma_1\partial \varsigma_2$ is a constant. 
Writing 
$b_1(u)=B_{22}/B_{21}$ and 
$b_2(u)=B_{12}/B_{11}$,
equations (\ref{Aijuu}) and (\ref{Aij}) give 
\begin{align}
\label{}
A_{11}=\frac{\mathsf a_{1}}{\sqrt{(b_1)_u}} \,, \qquad 
A_{21}=\frac{\mathsf a_{2}}{\sqrt{(b_2)_u}} \,, \qquad 
\{b_i(u), u \}_u =-2 a_0(u)\,,
\end{align}
where $\mathsf a_{1}$ and $\mathsf a_{2}$ are complex constants, and 
$\{b(u), u \} _u\equiv (2b_u b_{uuu}-3b_{uu}^2)/(2b_u^2)$ 
is the Schwarzian derivative with respect to $u$. 
Thus, the condition $\{b_1(u), u \} _u=\{b_2(u), u \} _u$ implies that they are related by
the M\"obius transformation
\begin{align}
\label{b2sol}
b_2(u)=\frac{\mathsf c_1 b_1(u)+\mathsf c_2}{\mathsf c_3 b_1(u)+\mathsf c_4}\,, \qquad 
\mathsf c_1\mathsf c_4-\mathsf c_2\mathsf c_3\ne 0\,,
\end{align}
where $\mathsf c_{1-4}$ are complex constants. These are all constraints coming from 
equations (\ref{chiz}) and (\ref{chiuu}).

To encapsulate, the Case (I) Petrov-III solution (\ref{Case1sol}) preserves half of supersymmetry, 
provided that  the base space is conformally spherical, viz, $P=2|\psi| P_*$ with $K=4|\psi|^2$, where 
$P_*$ and $\psi$ are  given by (\ref{Pstar}) and (\ref{psiCaseI}), respectively. 
The functions $f_i$ are given by 
\begin{align}
\label{Case1fi}
f_1(u,z)=B_{11}(u)\left[\varsigma_1(z)+b_2(u)\varsigma_2(z)\right]\,, \qquad 
f_2(u,z)=B_{21}(u)\left[\varsigma_1(z)+b_1(u)\varsigma_2(z)\right]\,, 
\end{align}
where $\partial^2 \varsigma_i=h_0(z)\varsigma_i$ and $b_2(u)$ is given by (\ref{b2sol}). 
$B_{11}$ and $B_{21}$ are subjected to 
\begin{align}
\label{Case1B21s}
B_{21}=\frac{W_0}{B_{11}(b_1-b_2)}\,, \qquad 
|B_{11}|^4=-\frac{2{\mathsf  a}_{2}^2|W_0|^2 (\bar b_1)_u}{\bar{\mathsf  a}_{1}^2 (b_2)_u (\bar b_1-\bar b_2)^2}\,.
\end{align}
The components of the Killing spinor are given by (\ref{epsilon32CaseI}) and (\ref{chi}), where 
$\varepsilon_i=\mathsf a_i(\varsigma_1+b_i(u)\varsigma_2)/\sqrt{(b_i)_u}$. 
It follows that functions $\psi$ and $P_*=P/(2|\psi|)$ appearing in the metric read
\begin{subequations}
\begin{align}
\label{}
\psi =&\,\frac{i(b_1)_u[\mathsf c_3\varsigma_1^2+(\mathsf c_1-\mathsf c_4)\varsigma_1\varsigma_2-\mathsf c_2\varsigma_2^2]}
{2g(\varsigma _1+b_2 \varsigma_2)[(\mathsf c_3b_1+\mathsf c_4)\varsigma _1+(\mathsf c_1 b_1+\mathsf c_2)\varsigma _2]}\,, 
 \\ 
P_*=&\,\frac{|B_{11}|^2|\varsigma _1+b_2\varsigma _2|^2+2|B_{21}|^2|\varsigma _1+b_1\varsigma _2|^2}{2|\mathcal W B_{11}B_{21}(b_1-b_2)|}\,.
\end{align}
\end{subequations}
Note that one can always normalize the Wronskians as $W_0=\ma W=1$ and thus the 
solution is controlled by two functions $b_1(u)$ and $h_0(z)$.

An important lesson here is that the first integrability conditions (\ref{intcond}) are not sufficient to guarantee the existence of a 
Killing spinor. The second integrability conditions $\mas R_{(\mu)}=0$ put 
more restrictions on the configuration of $P=P(u, z, \bar z)$.\footnote{
In hindsight, we can make prognostications about the nonexistence of the Killing spinor for generic
$P$ and $\psi$ from the outset, since, as it stands, the solution (\ref{Case1sol}) does not admit any timelike or null 
Killing vectors. On the other hand, the solution with (\ref{Case1fi}) and (\ref{Case1B21s}) allows a 
timelike Killing vector 
\[
\Lambda_r \partial_u -\Lambda_u\partial_r -2g\left(\psi \varepsilon_1\varepsilon_2\partial
+\bar\psi \bar\varepsilon_1\bar\varepsilon_2\bar\partial\right)\,, \qquad 
\Lambda\equiv \frac{r(|\varepsilon_1|^2+|\varepsilon_2|^2)}{\sqrt 2 P_*}\,,
\]
as a bilinear vector of the Killing spinor.
} 
The classification of spacetimes that admit Killing vectors follows the same line of reasoning; see \cite{Nozawa:2019dwu} for reference.

It is instructive to see the properties of the solution (\ref{Case1sol}) in more detail. 
Notwithstanding the absence of any curvature singularities $R_{\mu\nu\rho\sigma}R^{\mu\nu\rho\sigma}=24g^4$,  
the solution exhibits a p.p. curvature singularity at $r=0$. 
On top of this, the spacetime suffers from a different type of singular behavior. 
When the surface $S$ spanned by ($z, \bar z$) is compact and regular, the integration of 
$\Delta_2 K=8P^2 |\phi_2|^2$ over $S$ gives
\begin{align}
\label{}
8\int_S P^2 |\phi_2|^2 \D S=\int _S \Delta_2 K \D S=0 \,, 
\end{align}
leading to $\phi_2=0$, i.e., the spacetime is AdS. 
It follows that this Petrov-III radiating solution fails to describe a regular spacetime.

\subsection{Case (II): $Q=\bar Q(\ne 0)$}
\label{sec:realQ}

Let us next consider the Case (II) where $Q$ is real and nonvanishing, i.e., 
$q\ne 0$ and $\sin \alpha =0$. 
If $Q$ is real, $z$ derivative of $Q(u,z)=\bar Q(u, \bar z)$ implies 
$Q=Q(u)$, which can be taken to be a real constant $Q=\mathsf Q_e(\ne 0)$ by the reparametrization (\ref{repara}). 
This renders $X_2=X_4=Z_4=0$. It follows that $\ma B_1$ and $\ma B_4$ reduce to the sum of 
positive-definite terms, leading to $X_1=X_3=0 $ and $Z_1=Z_2=0$. 
We therefore have  $P=P(z,\bar z)$, $\phi_2=0$ and $\partial K=0$. 
Thus,  $\D s_2^2=2P^{-2}\D z \D \bar z$ is a space of  constant Gauss curvature $K(u,z,\bar z)=k={\rm constant}$ and 
$M^2=k\mathsf Q_e^2={\rm constant}$, leading to $k=0,1$. 
Now, the field equations are automatically satisfied. 

It follows that the solution is $u$-independent and recovers the Reissner-Nordstr\"om-AdS family
endowed with an electric charge
\begin{align}
\label{RN}
\D s^2=-2\D u \D r-\left[\left(\sqrt k-\frac{\mathsf Q_e}{r} \right)^2 +g^2 r^2\right]\D u^2+r^2 \D \Sigma_{k\ge 0}^2 \,, \qquad 
A=-\frac{\mathsf Q_e}{r}\D u \,,
\end{align}
where $\D \Sigma_k^2$ with $k=0,1 $ is given by (\ref{dSigmasq}). 
The supersymmetry of this solution was first demonstrated by Romans~\cite{Romans:1991nq}.
This solution admits more supersymmetry than minimally required. 

Note that this solution always admits a naked singularity at $r=0$, since 
it is not veiled by event horizon. 
This should be contrasted with the asymptotically flat case ($g=0$ and $k=1$), in which 
the solution describes a black hole with a degenerate event horizon.

\subsection{Case (III): $R_4=0$ ($Q\ne \bar Q$)} 

The Case (III) is defined by 
$R_4=0$ with $Q\ne \bar Q$, which leads to $Q=Q(u)$. 
The solution to (\ref{EOM2}) becomes $Q=Q_0(u) e^{i\mathsf a}$, 
where $Q_0(u) $  is a real function of $u$ and $\mathsf a$ is a real constant. 
By the reparametrization freedom (\ref{repara}), one can set $Q_0(u)=\mathsf Q$, 
where $\mathsf Q$ is a real constant. Thus, the electric and magnetic charges 
are constants 
\begin{align}
\label{}
\mathsf Q_e=\mathsf Q\cos\mathsf a \,, \qquad 
\mathsf Q_m=\mathsf Q\sin\mathsf a (\ne 0)\,,
\end{align}
where $\sin\mathsf a\ne 0$ follows from $Q\ne \bar Q$. 
Note that we cannot set $\mathsf a=0$ by the electromagnetic duality, 
since the duality invariance is broken for the Killing spinor equation (\ref{Killingspinor}). 
For the constant charge $Q=\mathsf Q e^{i \mathsf a}$, we have $X_4=Z_4=0$. Since we have assumed $\sin\mathsf a \ne 0$, 
equation (\ref{B1p}) implies that $R_1$ and $X_1$ do not vanish simultaneously, which enables us to solve  $\ma B_1=\ma B_2=0$ as  
\begin{align}
\label{III_Malpha}
M=\frac{\mathsf Q^2[X_1X_3-2R_1R_2\cos(\Theta_1-\Theta_2)]}{2(2R_1^2+X_1^2)}\,, \qquad 
\sin\mathsf a =\pm \frac{\sqrt{2R_1^2+X_1^2}}{2g\mathsf Q^3}\,,
\end{align} 
Substituting these conditions into $\ma B_4=0$, 
we obtain 
\begin{align}
\label{}
(2R_1^2+X_1^2)R_2^2 \sin^2(\Theta_1-\Theta_2)+\left[X_1R_2\cos(\Theta_1-\Theta_2)+R_1X_3\right]^2 =0 \,.
\end{align}
Since this equation is the sum of perfect squares and we have assumed $2R_1^2+X_1^2>0$
(by $\mathsf Q\sin\mathsf a \ne 0$), 
we need 
\begin{align}
\label{eqIII}
R_2 \sin(\Theta_1-\Theta_2)=0\,, \qquad  
X_1R_2\cos(\Theta_1-\Theta_2)+R_1X_3=0\,.
\end{align}
Thus, the Case (III) solutions are divided into (III-i) $R_2=0$ or (III-ii) $R_2\ne 0$. The Case (III-i) is further branched into 
subcategory as (III-i-a) $X_3=0$ and (III-i-b) $X_3\ne 0 $.

\subsubsection{Case (III-i-a): $R_2=X_3=0$} 

In this case, equation (\ref{III_Malpha}) implies $M=0$. 
The conditions $R_2=X_3=0$ are then boiled down to $\partial K=P_u=0$, 
implying that the two-dimensional space $\D s_2^2=2P^{-2}\D z\D \bar z$ must be a space 
of constant Gauss curvature $K=k=\{0, \pm 1\}$. 
Thus, we have $\phi_2=0$ by (\ref{EOM1}). 
Lastly, $\ma B_1=0$ puts the following constraint to the magnetic charge 
\begin{align}
\label{}
\mathsf Q_m =\pm \frac{k}{2g}\,.  
\end{align}
This is reminiscent of Dirac's quantization condition. 
Since Class (III) requires $Q\ne \bar Q$, we need $k\ne 0$. 
It follows that the solution  is static and takes the form
\begin{align}
\label{cosmicdyon}
\D s^2=&\,-2 \D u \D r-\left[\left(\frac{k}{2gr}+gr\right)^2+\frac{\mathsf Q_e^2}{r^2}\right]
\D u^2+r^2 \D \Sigma_{k=\pm 1}^2 \,, 
\\
A=&\,-\frac{\mathsf Q_e}{r}\D u\mp  \frac{i}{2g}\left[\partial(\ln P_{k})\D z-\bar \partial (\ln P_{k})\D \bar z\right]\,.
\end{align}
where $P_k$  ($k=\pm 1$) is given by (\ref{Pk}). Note that the value of electric charge $\mathsf Q_e$ remains unrestricted under supersymmetry.
The Killing spinor for this solution was already constructed by \cite{Romans:1991nq}. The solution keeps one quarter of supersymmetry. 
This solution illustrates the asymmetry between electric and magnetic charges in the gauged case.

This solution is referred to as the cosmic dyon in \cite{Romans:1991nq}, which 
does not allow the flat space counterpart, i.e., the $g\to 0$ limit cannot be taken.
The solution asymptotically approaches to AdS with $\mathsf Q_m\ne 0$, which defines a vacuum topologically 
distinct to (\ref{AdSmetric}) \cite{Hristov:2011ye}.  Indeed, this magnetic AdS vacuum is not maximally supersymmetric.  
When $\mathsf Q_e\ne 0$ or $k=1$, the singularity at $r=0$ is naked. 
The unique option to evade the naked singularity is $k=-1$ and $\mathsf Q_e=0$.

\subsubsection{Case (III-i-b): $R_2=0$ and $X_3\ne 0$}  

From (\ref{eqIII}), we have $R_1=0$ with $X_1\ne 0$. 
Since Case (III) is characterized by $Q(u,z)=\mathsf Qe^{i \mathsf a}={\rm constant}$,  $R_1=0 $ implies $\phi_2=0$, while 
equations of motion (\ref{EOM1}) and (\ref{EOM4}) lead to $M=M(u)$ and $P_u=0$. 
The last condition is incompatible with the assumption $X_3\ne 0$ for the 
Case (III-i-b). We therefore have  no supersymmetric solutions belonging to Case (III-i-b).

\subsubsection{Case (III-ii): $R_2\ne 0$} 

In this case, the first condition in (\ref{eqIII}) reduces to $\sin(\Theta_1-\Theta_2)=0$, 
viz, $\Theta_2=\Theta_1-\frac 12(1+\delta_{1}^{ \pm})\pi$ with $\delta_{1}^{\pm}=\pm 1$. 
This implies that $Z_2/Z_1=-\delta_1^\pm r_2/r_1$. 
Substitution of this  into the second equation of (\ref{eqIII}) yields
\begin{align}
\label{Kz_III-ii}
\partial K=-\frac{2 e^{-i\mathsf a}\phi_2}{\mathsf Q(\mathsf Q^2 K-M^2)P}\left[M(\mathsf Q^2K-M^2)P+4\mathsf Q^4 P_u\right]\,.
\end{align}
Inserting this into $\ma B_2=0$, we have 
\begin{align}
\label{Pu_III-ii}
P_u=\frac{2g^2\mathsf Q^2 M P \sin^2\mathsf a (\mathsf Q^2K-M^2)}{(\mathsf Q^2K-M^2)^2+8\mathsf Q^4 P^2 |\phi_2|^2} \,. 
\end{align}
Substituting this back into (\ref{Kz_III-ii})
and taking its complex conjugation,  
the integrability condition $(\partial \bar\partial -\bar\partial \partial)K=0$ gives rise to 
\begin{align}
\label{}
M\left[\partial M \bar \partial (P^2 |\phi_2|^2)-\bar \partial M\partial (P^2 |\phi_2|^2)\right]=0\,.
\end{align}
Supposing  $M=0$, equations (\ref{Kz_III-ii}) and (\ref{Pu_III-ii}) imply $P_u=\partial K=0$, which 
undermines the assumption $R_2\ne 0$ of Case (III-ii). The $\phi_2=0$ case  is also excluded by $R_2\ne 0$. 
It follows that $P$ takes the following form
\begin{align}
\label{Ph_III-ii}
P(u, z, \bar z)=\frac{\hat P(u, M(u,z,\bar z))}{\sqrt{\phi_2(u,z,\bar z)\bar \phi_2(u,z,\bar z)}} \,,
\end{align}
where $\hat P$ is real. 
Plugging this into $\ma B_1=0$, 
we get 
\begin{align}
\label{K_III-ii}
K=\frac{M^2}{\mathsf Q^2}+2\delta_{2}^\pm\sqrt{g^2\mathsf Q^2 \sin^2\mathsf a -2 \hat P^2} \,, 
\qquad \delta_{2}^\pm=\pm 1 \,.
\end{align}
Replacing the right-hand side of (\ref{Kz_III-ii}) by 
(\ref{Ph_III-ii}) and (\ref{K_III-ii}),
we obtain 
\begin{align}
\label{Kz2_III-ii}
\partial K=-\frac{6 e^{-i\mathsf a} M \phi_2}{\mathsf Q} \,. 
\end{align}
The consistency of (\ref{K_III-ii}) and (\ref{Kz2_III-ii}) brings about 
\begin{align}
\label{hatP_III-ii}
\hat P(u, M)=\sqrt{
\frac 12 g^2 \mathsf Q^2 \sin^2\mathsf a -\frac 1{32\mathsf Q^4}\left(M^2-4\mathsf Q^4 P_0(u)\right)^2
}\,,
\end{align}
where $P_0$ is an arbitrary function of $u$ and $\delta_2^\pm ={\rm sgn}(M^2-4\mathsf Q^4P_0(u))$. 
We are thus led to 
\begin{align}
\label{Ks_III-ii}
K(u, z, \bar z)=\frac{3M(u,z,\bar z)^2-4\mathsf Q^4 P_0(u)}{2\mathsf Q^2}\,.
\end{align}
Acting $\bar\partial $ to (\ref{Kz2_III-ii}) and 
eliminating $\Delta_2 K$  from (\ref{EOM3}),  
we obtain
\begin{align}
\label{Mu_III-ii}
M_u=\frac{16g^2 \mathsf Q^6\sin^2\alpha-[M^2-4\mathsf Q^4 P_0(u)]^2}{8\mathsf Q^4}\,.   
\end{align}
Using (\ref{EOM1}), the compatibility $\partial M_u-(\partial M)_u=0$ 
implies 
\begin{align}
\label{phi2u_III-ii}
(\phi_2 )_u =-\frac{M(M^2-4\mathsf Q^4 P_0(u))}{2\mathsf Q^4}\phi_2 \,. 
\end{align}
Inserting (\ref{Ph_III-ii}), (\ref{hatP_III-ii}), (\ref{Ks_III-ii}), (\ref{Mu_III-ii}) and (\ref{phi2u_III-ii})
 into (\ref{Pu_III-ii}), we obtain 
 \begin{align}
\label{}
\big(M^2-4\mathsf Q^4 P_0(u)\big) P_0'(u)=0\,. 
\end{align}
For the  $M=\pm 2 Q^2 \sqrt{P_0(u)}$ case, 
equation (\ref{EOM1}) is not satisfied since we have assumed $\bar Q\phi_2\ne 0$. Thus we have 
 $P_0(u)= \mathsf C$, where $\mathsf C$ is a real constant.
 One can then verify that 
 $[(M^2-4\mathsf Q^4 \mathsf C)^2-16g^2\mathsf Q^6\sin^2\mathsf a]/\bar \phi_2$ is $z$-independent, 
 implying a nonvanishing function $h=h(u,z) $ such that 
 \begin{align}
\label{phi2_III-ii}
\phi_2(u,z,\bar z)=h(u,z) \left[(M^2-4\mathsf Q^4 \mathsf C)^2-16 g^2\mathsf Q^6\sin^2\mathsf a \right]\,. 
\end{align}
The compatibility of (\ref{phi2u_III-ii}) and (\ref{phi2_III-ii}) implies $h=h(z)$. 
By the reparametrization (\ref{repara}), one can always set $h(z)=e^{i\mathsf a }/(16\mathsf Q^4)={\rm constant}$. 
With this gauge fixing, $M$ obeys 
\begin{align}
\label{}
\partial M-\mathsf Q M_u=0 \,,
\end{align}
by virtue of (\ref{EOM1}), (\ref{Mu_III-ii}) and (\ref{phi2_III-ii}). 
Since $M$ is real, it turns out that $M$ depends on a 
single variable $\eta=u+\mathsf Q(z+\bar z)$ as $M=M(\eta)$, which is constrained by (\ref{Mu_III-ii}) as 
\begin{align}
\label{dMdeta}
\frac{\D M}{\D \eta}= 2g^2\mathsf Q^2\sin^2\mathsf a-\frac{(M^2(\eta)-4\mathsf C\mathsf Q^4)^2}{8\mathsf Q^4}\,,
\end{align}
giving
\begin{align}
\label{}
 P=\left[2g^2\mathsf Q^2\sin^2\mathsf a-\frac{(M^2-4\mathsf C\mathsf Q^4)^2}{8\mathsf Q^4}\right]^{-1/2}\,, \qquad 
 K=\frac{3M^2-4\mathsf C \mathsf Q^4}{2\mathsf Q^2}\,.
\end{align}

We have exhausted all the constraints coming from supersymmetry conditions and equations of motion. In order to illustrate the physical interpretation of spacetime, we are now going to switch the metric into a more familiar set of coordinates.  To this end, it is of convenience to 
employ $M$ itself as an independent variable as $x=M/(2\mathsf A\mathsf Q^2)$, where 
$\mathsf A$ is a constant. 
Performing the coordinate transformation
\begin{align}
\label{}
z=\frac{\mathsf A(\eta-u)+i\varphi}{2\mathsf A\mathsf Q}\,, \qquad 
r=\frac{1}{\mathsf A(x-y)} \,, \qquad \D t=\mathsf A \D u+\frac{\D y}{\Delta_y(y)}\,, 
\end{align}
with 
$\D \eta=\mathsf A \D x/[g^2 \sin^2\mathsf a-\mathsf Q^2(\mathsf C-\mathsf A^2x^2)^2]$
[c.f (\ref{dMdeta})], 
one arrives at the manifestly static form 
\begin{align}
\label{Cmetric}
\D s^2=&\,\frac{1}{\mathsf A^2(x-y)^2}\left(-\Delta_y(y)\D t^2+\frac{\D y^2}{\Delta_y(y)}+\frac{\D x^2}{\Delta_x(x)}+
\Delta_x(x)\D \varphi^2 \right)\,, \\
A=&\, \mathsf Q \cos\mathsf a y \D t+\mathsf Q \sin\mathsf a x \D \varphi \,, 
\end{align}
where we have dropped the exact form from $A$, and 
the structure functions are given by 
\begin{align}
\label{Deltaxy}
\Delta_y(y)=\frac{g^2}{\mathsf A^2}\cos^2\mathsf a+\frac{\mathsf Q^2}{\mathsf A^2}\left(\mathsf C-\mathsf A^2 y^2\right) ^2 \,, \qquad 
\Delta_x(x)=\frac{g^2}{\mathsf A^2}-\Delta_y(x) \,. 
\end{align}
This spacetime is commonly referred to as the C-metric and belongs to the Petrov-type D. 
The supersymmetry and the existence of Killing spinor was addressed in 
\cite{Klemm:2013eca}. 

 The C-metric in the Einstein-$\Lambda$ system describes either a pair of accelerated black holes 
 or an accelerated black hole in AdS, depending on the parameters. See e.g., \cite{Podolsky:2002nk,Dias:2002mi} for details. 
Remark that the supersymmetric C-metric does not  permit the Killing horizon when the electric charge is nonvanishing, since 
$\Delta_y(y)$ is expressed as a sum of squares, for which the curvature singularity at $y=-\infty $ is visible from an observer at infinity 
$x=y$. The degenerate horizon exists only for the purely magnetic case $\cos \mathsf a=0$. 
The requirement of the nonvanishing magnetic charge ($\sin \mathsf a\ne 0$) is also apparent
from the Lorentzian condition
$\Delta_x(x)=[g^2\sin^2\mathsf a -\mathsf Q^2(\mathsf C-\mathsf A^2 x^2)^2]/\mathsf A^2>0$.

\subsection{Case (IV): $R_1=0$ ($R_4\ne 0$)} 

The Case (IV) corresponds to $(Q-\bar Q)R_4\ne 0$ and $R_1=0$. Solving $Z_1=0$, we have 
\begin{align}
\label{phi2IV}
\phi_2= -\frac{M\partial Q}{2|Q|^2} \,.
\end{align}
The constraint  (\ref{constraints2}) implies $X_4=0$, which is solved as
 $Q(u, z)=Q_0(u)Q_1(z)(\ne 0)$, where $Q_0$ is a real function of $u$ and 
$Q_1$ is holomorphic in $z$. 
By the reparametrization (\ref{repara}), one can set $Q_0=1$ without loss of generality. 
The  condition $R_4\ne 0 $ requires $Q_1(z)$ not to be constant.
Equations (\ref{EOM1}) and (\ref{phi2IV}) yield $M=M_0(u)Q_1(z)\bar Q_1(\bar z)$, where $M_0$ is a real function of $u$. 
Plugging this into (\ref{EOM4}), we obtain $P_u=0$, resulting in $X_3=Z_3=0$. Equation 
$\ma B_2=0$ gives $M_0=0$ and hence $\phi_2=0$ due to (\ref{phi2IV}), whereas 
equation $\ma B_1=0$ derives
\begin{align}
\label{}
K(z,\bar z)=\pm i g \Big(Q_1(z)-\bar Q_1(\bar z)\Big)\,.
\end{align}
Other equations $\ma B_3=\ma B_4=\ma B_5=0$ and equations of motion are trivially satisfied.  
The solution is then static and reads ($\partial Q_1\ne 0$)
\begin{subequations}
\label{moreon}
\begin{align}
\D s^2=&\,-2 \D u \D r- 
\left|\pm\frac{iQ_1(z)}{r}+g r\right|^2\D u^2
+\frac{2r^2}{P(z,\bar z)^2}\D z\D \bar z \,, 
\\
A=&\,-\frac{Q_1(z)+\bar Q_1(\bar z)}{2r}\D u\pm \frac{i}{2g}\left(\partial (\ln P)\D z-\bar 
\partial (\ln P)\D \bar z\right)\,,
\end{align}
\end{subequations}
where $P=P(z,\bar z)$ satisfies $K\equiv 2P^2 \partial\bar \partial \ln P=\pm i g (Q_1-\bar Q_1)$, i.e., $\Delta_2 K=0$. 
No other restrictions are placed upon $P$. This is a solution obtained in \cite{Cacciatori:2004rt} and indeed admits 
a Killing spinor. 

The charge function $Q_1(z)$ is pertinent to the topology of  the angular surface 
$S$ spanned by $(z, \bar z)$ as follows. Assuming $S$ is compact and regular, the Gauss-Bonnet theorem implies
\begin{align}
\label{}
\chi=\frac 1{4\pi}\int_S K \D S =\pm \frac{ig}{4\pi}\int _S\big(Q_1(z)-\bar Q_1(\bar z)\big)\D S \,,
\end{align}
where $\chi$ is the Euler characteristic of $S$. For any value of $Q_1$, 
the present solution possesses a naked singularity at $r=0$, 
on account of $g_{uu}>0$.

\subsection{Case (V): $\cos(\alpha+\Theta_1+\Theta_4)=0$ ($R_1R_4\ne 0$)}

Finally, we consider the case (V) where $\cos(\alpha+\Theta_1+\Theta_4)=0$ ($R_1R_4\ne 0$). 
As we will demonstrate below, it turns out that this case does not admit supersymmetric solutions. 
Given the extensive mathematical intricacies and calculations involved, 
readers uninterested in these details may opt to bypass this section.

For $R_1R_4\ne 0$, equation (\ref{B3p}) requires $\cos(\alpha+\Theta_1+\Theta_4)=0$, i.e., 
$\Theta_4=\pm \pi/2-(\alpha+\Theta_1)$. 
Eliminating $X_4$ from  (\ref{constraints2}), 
we obtain 
\begin{align}
\label{calB5V}
\ma B_5=\pm \frac{2 R_4}{Q}
\left[2MR_1 \sin(2\alpha)+Q^2 R_2 \sin(2\alpha+\Theta_1-\Theta_2)\right]=0 \,. 
\end{align}
Assuming $\sin(2\alpha)=0$ ($\alpha=\pm \pi/2$), we have $Q(u,z)=-\bar Q(u,\bar z)$. 
This leads to $\partial Q=0$, which is inconsistent with $R_4 \ne 0$. Thus we find that
$Q+\bar Q=2i q \cos\alpha$ is nonvanishing. 
It is also concluded that $X_4\ne 0$, since equation (\ref{constp}) implies 
$gq^2 X_4=2 R_1 R_4 \sin(\Theta_1+\Theta_4)=2R_1R_4\cos\alpha \ne 0$. 
Assuming next $M=0$, we have $\phi_2=0$ by (\ref{EOM1}), giving rise to $Z_1=0$. 
This also runs counter to the assumption $R_1\ne 0$ of Case (V). 
Owing to $MR_1\sin(2\alpha )\ne 0$, we need 
$R_2\sin(2\alpha +\Theta_1+\Theta_2)\ne 0$ by (\ref{calB5V}). 
The condition $Z_4\ne 0$ also implies the electromagnetic 
phase $\alpha$ fails to be constant, since constant $\alpha$ implies 
$Q=Q(u)$, which does not comply with the assumption $R_4\ne 0$ of Case (V). 
Thus, we allege $MR_1R_4 R_2X_4 \D \alpha \ne 0$ 
in the following analysis. 

Combining  $\ma B_3=0$ and (\ref{EOM2}), we obtain
\begin{align}
\label{phi2V}
Z_1=-\frac{i g Q\bar Q^2X_4}{(Q+\bar Q)Z_4} \,. 
\end{align}
Inserting this into $\ma B_2=\ma B_5=0$, we find
\begin{align}
Z_2=-\frac{2M\bar QX_2X_4}{(Q^2+\bar Q^2)Z_4}+\frac{i(Q+\bar Q)(2M |Q|^2X_2^2-X_1X_3)\bar Z_4}{g\bar Q(Q^2+\bar Q^2)X_4}\,.
\label{KzV}
\end{align}
Note that $Q^2+\bar Q^2\ne 0$ due to $\alpha\ne {\rm constant}$. 
Next,  we define real functions 
\begin{align}
\label{}
Y\equiv -g^2(Q-\bar Q)^2-\frac{2|Q|^2X_4^2}{P^2|\partial Q|^2(Q+\bar Q)^2}\,, \qquad 
X_\pm \equiv \frac{i Q^{1-\delta_1^\pm}\bar Q^{1-\delta_1^\pm}(Q^{2 }-\bar Q^{2})^{\delta_1^\pm}}{P^4} \,, 
\end{align}
where 
$\delta_1^\pm=\pm 1$. Here, $X_+$ corresponds to $\delta_1^\pm =1$, and $X_-$ corresponds to $\delta_1^\pm =-1$. 
The nonnegativity of $Y$ follows from $\ma B_1=X_1^2-|Q|^2 Y=0$. 
Supposing $Y=0$, we have $X_1=M^2-K |Q|^2=0$. By inserting $K=M^2/|Q|^2$ into (\ref{KzV}) and its complex conjugation, 
we end up with a contradiction $X_4=0$, yielding no solutions for $Y=0$.  For $Y>0$, 
we can solve $\ma B_1=\ma B_4=0$ with (\ref{phi2V}) and (\ref{KzV}) as
\begin{align}
\label{MV}
M=&\,\frac{\delta_2^\pm |Q|^2}{2\sqrt{Y}}(\ln X_\pm)_u \,, \\
\label{KV}
K=&\,\frac{M^2}{|Q|^2}-\delta_2^\pm \sqrt{Y}\,, 
\end{align}
where $\delta_2^\pm=\pm 1$. The emergence of four branches of $M$ reflects the fact that $\ma B_1$ is quartic in $M$.
The relation  (\ref{KzV}) is then reduced to 
\begin{align}
\label{KzVs}
\partial K=\frac{M^2\partial Q}{Q^2\bar Q}+\delta_1^\pm \delta_2^\pm \frac{\sqrt{Y}\partial Q}{Q-\bar Q}
+\frac{3i M\bar Q X_4}{P^2 Q(Q+\bar Q)\bar\partial \bar Q}\,.
\end{align}
Using (\ref{EOM1}) and (\ref{phi2V}), the compatibility of (\ref{KV}) and (\ref{KzVs}) gives 
\begin{align}
\label{PzV}
\frac{X_4^2}{(\partial Q)^2}\partial \left(\frac{\partial Q}{X_4^2}P^2\right)
=&\, -i\frac{P^4 \bar Q(Q+\bar Q)}{Q^2X_4}\left(\frac{Q}{P^2}\right)_u
+\frac{P^2(\bar Q^2+4|Q|^2-Q^2)}{2Q^2(Q+\bar Q)}+\delta_1^\pm \frac{P^2(3Q^2-\bar Q^2)}{2Q^2(Q-\bar Q)}
\notag \\
&\,+g^2 (1+\delta_1^\pm)\frac{P^4(Q-\bar Q)(Q+\bar Q)^2|\partial Q|^2}{|Q|^2X_4^2} \,.
\end{align}
Inserting (\ref{phi2V}) into (\ref{EOM4}) and eliminating $\partial P$ by (\ref{PzV}), 
we obtain $\ma O_1=0$, where 
\begin{align}
\ma O_1 \equiv &\, i X_4 Q^3 (Q^2-\bar Q^2)^2 \left[\partial \ln \left(\frac{i(Q^2-\bar Q^2)}{|Q|^4}\right)\right]_u
\notag \\
&+(1+\delta_1^\pm)(Q+\bar Q)\partial Q\left[2P^2 Q|\partial Q|^2(Q^2-\bar Q^2)^2g^2+(3Q^2-\bar Q^2)\bar QX_4^2\right]\,.
\label{eqQV}
\end{align}
If we take $\delta_1^\pm=-1$ in (\ref{eqQV}), we obtain
$i (Q^{-2}-\bar Q^{-2})=h_0(u)h_1(z)\bar h_1(\bar z)$, where $h_0$ is a real function of $u $ and $h_1$ is a holomorphic function.
Taking the $z$ derivative of this equation,  we find that $h_1(z)$ needs to be constant. 
This requires $Q=Q(u)$, leading to the contradiction to the assumption of Case~(V). 

We shall next consider the case $\delta_1^\pm=+1$. Computing 
the integrability condition $(\partial\bar\partial-\bar\partial\partial )K=0$ from (\ref{KzV}) and using $\ma O_1=\ma{\bar O}_1=0$, 
we find another obstruction  $\ma O_2 =0$, where 
\begin{align}
\label{}
\ma O_2 \equiv & \left(\frac{Q^2-\bar Q^2}{P^4}\right)_u \,. 
\end{align}
Solving $\ma O_2=0$, we find $P=[i (Q^2-\bar Q^2)]^{1/4}P_1(z,\bar z)$, where $P_1$ is real and 
independent of $u$. This profile of $P$ corresponds to $(X_+)_u=0$, and hence 
$M=0$ by (\ref{MV}). This violates the assumption of Case (V).

In conclusion, we do not have supersymmetric solutions in Case (V).

\section{Final remarks}
\label{sec:final}

In this paper, we  investigated the supersymmetry conditions for the 
Robinson-Trautman family of solutions in the Einstein-Maxwell theory with a negative cosmological constant. 
Making full use of integrability conditions (\ref{Bi}) of the Killing spinor equation, 
we comprehensively categorized all conceivable supersymmetric solutions. 
By a close investigation of supersymmetry conditions, we were able to find the explicit metric expressions. 
It turns out that five distinct classes of solutions
are realized; 
(I) the Petrov-III radiating solution (\ref{Case1sol}) with (\ref{Pstar}), (\ref{psiCaseI}) and (\ref{varepsilonex})--(\ref{Case1B21s}), 
(II) the electric Reissner-Nordst\"om-AdS solution (\ref{RN}), (III-i-a) the cosmic dyon (\ref{cosmicdyon}), 
(III-ii) the C-metric (\ref{Cmetric}) and (IV) the Cacciatori-Caldarelli-Klemm-Mansi (CCKM) solution (\ref{moreon}). 
The supersymmetry for Case (I) is new, while other solutions have already been shown to be supersymmetric in the literature. 
The significance of the present work lies in demonstrating  that these are exhaustive. 
Apart from Case (I), the solution is static. In either case, the solution describes a naked singularity unless some parameters are tuned. 
Physical properties are summarized in table~\ref{TableRT}.

\begin{table}
  \centering 
\begin{tabular}{c|c|c|c|c|c|c}
   & metric &   SUSY & $g\to 0$  &horizon   & static   & Petrov-type  \\
\hline\hline
Case (I) & (\ref{Case1sol}) & $1/2$ & no & no & no & III \\
Case (II) electric RN-AdS & (\ref{RN}) & $1/2$ &$\checkmark$ & $g=0$, $k=1$ &  $\checkmark$ & D \\
Case (III-i-a) cosmic dyon & (\ref{cosmicdyon}) & $1/4$ & no & $\mathsf Q_e=0$, $k=-1$ & $\checkmark$  & D \\
Case (III-ii) C-metric & (\ref{Cmetric}) & $1/4$ & no & $\mathsf Q_e=0$ & $\checkmark$ & D \\
Case (IV)  CCKM solution & (\ref{moreon}) & $1/4$ & no & no & $\checkmark$ & II \\
\hline
\end{tabular}
  \caption{Summary of the supersymmetric Robinson-Trautman solution.}
  \label{TableRT}
\end{table}

We highlighted that the electromagnetic duality invariance is broken for the Killing spinor equation. Obviously, 
the electric and magnetic charges are not on equal footing for these supersymmetric configurations. 
To restore the electromagnetic duality invariance, the analysis needs to be discussed within the symplectically invariant framework of supergravity~\cite{deWit:2011gk}. This is an interesting avenue, but beyond the central scope of this paper.

This paper concentrated on the aligned case, in which one of 
the principal null directions of the electromagnetic field is parallel to the principal null direction of the Weyl tensor. 
Since the non-aligned case is more complex, the possible metric expression is not yet obtained in a closed form,
see \cite{VandenBergh:2020lvf} for the study of Petrov-D. An accessible strategy to this problem is to
reconstruct the metric by imposing supersymmetry, as we have done in this paper.

In the present paper, attention was focused exclusively on the minimal model of gauged supergravity, for which the 
single AdS vacuum is realized by a pure cosmological constant. In gauged supergravity, various scalar fields belonging to vector and hypermultiplets 
may contribute to the bosonic Lagrangian. The critical points of scalar fields typically correspond to AdS vacua. 
If we allow the scalar fields to flow, regular geometries might be obtained in the supersymmetric limit. The uncharged Robinson-Trautman solution with scalar hair in supergravity 
has been recently constructed in \cite{Nozawa:2023boa}, which includes the hairy black hole~\cite{Faedo:2015jqa,Nozawa:2020gzz} and the C-metric \cite{Lu:2014sza,Nozawa:2022upa} as special cases. It has been  
demonstrated that the generalized Robinson-Trautman equation is tantamount to the integrability condition of the Ricci flow equation. It is thus of great interest to explore the relevance of the flow equation to supersymmetry. 
These issues are currently under investigation.

\acknowledgments

The author would like to thank Silke Klemm and Norihiro Tanahashi for stimulating discussions at an early stage of this work. 
The work of MN is partially supported by MEXT KAKENHI Grant-in-Aid for Transformative Research Areas (A) through the ``Extreme Universe'' collaboration 21H05189 and JSPS Grant-Aid for Scientific Research (20K03929).

\appendix 

\section{Ungauged case $g=0$}
\label{sec:g=0}

This appendix collects some issues that are valid
only for the ungauged case $g=0$.

\subsection{Supersymmetry}

Setting $g=0$ in (\ref{BiZX}), one finds
$X_1=Z_1=Z_2=0$ and $Z_3=(X_3-iX_4)/2=0$. 
Due to $X_4=0$, we find $Q=Q_0(u)Q_1(z)$, where $Q_0$ is real and therefore 
can be set to unity by the reparametrization (\ref{repara}).  
The condition $X_3=0$ then gives $P_u=0$, while 
equation (\ref{EOM4}) implies $\phi_2=\phi_2(u, z)$. 
The following analysis is split into two, according to 
$Q\ne 0$ or $Q=0$.

\subsubsection{$Q\ne 0$}

Eliminating $\phi_2$ from $Z_1=0$ and $Z_2=0$, and using $X_1=0$, we find 
$\partial K/K=\partial Q_1(z)/Q_1(z)$. Integration of this equation 
yields $K(z,\bar z)=CQ_1(z)\bar Q_1(\bar z)$, where the constancy of $C$ follows from 
$P_u=0$. This  implies $P=\sqrt{Q_1(z)\bar Q_1(\bar z)} P_0(z, \bar z)$ with $2P_0^2\partial \bar \partial \ln P_0$ 
being constant. 
By the reparametrization $z\to \zeta (z)$, one can set $Q_1$ to be constant, 
i.e., $K$ turns out to be constant $K=k=\{0, +1\}$. Here we note that $k=-1$ is excluded by
 $Z_1=0$. Equation   (\ref{EOM3}) then gives $\phi_2=0$, 
 which renders $M(u,z,\bar z)=\mathsf M$ and 
 to be constant by (\ref{EOM1}) and $Z_1=0$ and subjected 
to the  constraint $\mathsf M^2=k\mathsf Q^2$ in view of $X_1=0$.  
 The solution therefore reads
\begin{subequations}
\label{IWP}
 \begin{align}
\D s^2=&\,-2 \D u \D r -\left(k-\frac{\mathsf Q}{r}\right)^2 \D u^2 +r^2 \D \Sigma_{k\ge 0}^2 \,,
\\
A=&\, -\frac{\mathsf Q\cos \mathsf a}{r}\D u+i \frac{\mathsf Q\sin\mathsf a }{k}\left[\partial(\ln P_k)\D z-\bar \partial (\ln P_k)\D \bar z\right] \,.
\end{align}
\end{subequations}
This is the extreme Reissner-Nordstr\"om family contained in the most general 
Israel-Wilson-Perjes class~\cite{Perjes:1971gv,Israel:1972vx} preserving half of supersymmetry. 
As opposed to the $g\ne 0$ case, this solution admits an event horizon for $k=1$. 
It is also noteworthy to remark that one can apply the electromagnetic duality transformation 
to the ungauged solution (\ref{IWP}) to achieve, say, $\sin \mathsf a=0$, without  breaking supersymmetry, 
since the Killing spinor has no longer gauge charge for $g=0$.

 \subsubsection{$Q=0$}
 
 In the $Q=0$ case, we recover the metric (\ref{Case1sol}). 
 The subsequent argument is identical  up to (\ref{chi}). 
 Plugging (\ref{chi}) into (\ref{KSQ0u}), we obtain
 \begin{align}
\label{}
\psi(u,z)=\frac{\psi_1(z)}{\varepsilon_2(u,z)^2} \,, \qquad 
\bar\psi(u,\bar z)=\frac{\bar\psi_2(\bar z)}{\bar\varepsilon_1(u,\bar z)^2} \,,
\end{align}
where $\psi_i(z)$ are arbitrary holomorphic functions and 
$\varepsilon_i(u,z)$ satisfy $\partial^2 \varepsilon_i=h_0(z)\varepsilon_i$ and (\ref{chiz}). 
Then we obtain $P=2|\psi|P_*=2\sqrt{\psi_1(z)\bar \psi_2(\bar z)}P_*/(\bar\varepsilon_1\varepsilon_2)$.
 By the reparametrization (\ref{repara}) and $\partial^2 \varepsilon_i=h_0(z)\varepsilon_i$, one can set $\psi_1(z)=\bar \psi_2=1$, yielding
$\varepsilon_1=\pm\varepsilon_2$. This renders the independent components of the Killing spinor
down to one. However, the one quarter of supersymmetry is not allowed in the ungauged case~\cite{Tod:1983pm}. Thus, we conclude that the $Q=g=0$ Robinson-Trautman
solution does not admit the supersymmetric limit.

\subsection{Minkowski spacetime in the Robinson-Trautman form}

Setting $M=Q=\phi_2=g=0$ in (\ref{metric}), all components of the Riemann tensor vanish, 
leading to the Minkowski spacetime
\begin{align}
\label{Minkowski}
\D s^2=-\left[k-2 r \left(\ln P\right)_u\right]\D u^2-2 \D u \D r+\frac{2r^2}{P^2}\D z \D \bar z\,, 
\end{align}
where $P=P(u,z, \bar z)$ is given by
\begin{align}
\label{}
P=\frac{1+\frac 12 k |h_1|^2}{|\partial h_1|}\,, \qquad 
h_1(u, z)=\frac{f_1(u,z)}{f_2(u,z)}\,, \qquad 
\partial^2 f_i=h_0(z) f_i \,. 
\end{align}
To bring the above metric (\ref{Minkowski}) into the standard form, 
we define $u$-independent holomorphic functions $\varsigma_i (z)$ ($i=1,2$)
satisfying $\partial^2 \varsigma_i=h_0(z) \varsigma_i$. 
Since $\varsigma_i$ obey the second-order linear differential equations, 
their Wronskian $\ma W\equiv \varsigma_2 \partial\varsigma_1-\varsigma_1\partial \varsigma_2$ is constant, 
which can be set to $\ma W=1$ without loss of generality. 
In terms of these variables, we introduce new coordinates
\begin{align}
\label{}
\hat u= \frac{r|\varsigma_1|^2}{P}+\hat u_0(u) \,, \qquad 
\hat v=\frac{r|\varsigma_2|^2}{P}+\hat v_0(u)\,, \qquad 
\hat z=\frac{r\bar \varsigma_1 \varsigma_2}{P}+\hat z_0(u) \,, 
\end{align}
where 
\begin{align}
\label{}
\hat u_0(u)=&\, \int\left[P^2 \partial\bar \partial \left(P^{-1}|\varsigma_1|^2\right)+\frac{k|\varsigma_1|^2}{P}\right]\D u \,, \notag \\ 
\hat v_0(u)=&\,\int\left[P^2 \partial\bar \partial \left(P^{-1}|\varsigma_2|^2\right)+\frac{k|\varsigma_2|^2}{P}\right]\D u\,,\\
\hat z_0(u) =&\,\int \left[P^2 \partial\bar \partial \left(P^{-1}\bar \varsigma_1 \varsigma_2\right)+\frac{k\bar \varsigma_1\varsigma_2}{P}\right]\D u\,. 
\notag 
\end{align}
The equation $\partial^2 \varsigma_i=h_0(z) \varsigma_i$ assures 
that $\hat u_0$, $\hat v_0$ and $\hat z_0$ are independent of $z$ and $\bar z$. 
It is then straightforward to verify that the metric (\ref{Minkowski}) is rewritten into a familiar form
\begin{align}
\label{}
\D s^2=-2 \D \hat u \D \hat v+2 \D \hat z \D \hat{\bar z}\,. 
\end{align}

Although the standard metric of AdS is obtained in a similar fashion, 
we shall not attempt to do this here, 
since the expression is rather lengthy and not illuminating.

\end{document}